%% file: shapley.tex
\newif\ifmakeplots
\makeplotstrue

\newif\ifarxiv
\arxivfalse

\newif\ifjournal
\journalfalse

\ifjournal
    \documentclass[5p, times]{elsarticle}
    \usepackage{natbib}
    
\else
    \documentclass[12pt]{article}
    \usepackage[margin=1.25in, headheight=42pt]{geometry}
    \usepackage{hyperref}
    \usepackage{fancyhdr}
    \usepackage[style=alphabetic,backend=bibtex,giveninits=true,url=false,doi=false,
                natbib=true,maxbibnames=99]{biblatex}
\fi

\ifarxiv
    \newcommand{\datadir}{.}
\else
    \newcommand{\datadir}{data}
\fi

\usepackage[margin=1.25in]{geometry}
\usepackage{amsmath,amsthm,thmtools}
\usepackage{booktabs}
\usepackage{float}
\usepackage{cleveref}
\usepackage{url}
\usepackage{graphicx}
\usepackage{pgfplots}
\usepackage{algorithm2e}
\usepackage{environ}
\usepackage{filecontents}
\usepackage{subcaption}
\usepackage[font=small,labelfont=bf]{caption}
\usepackage{dcolumn}
\usepackage{tikz,tikz-3dplot}
\usepackage{multirow}

\newcolumntype{d}{D{.}{.}{-1}}

\theoremstyle{plain}

\usetikzlibrary{calc,arrows.meta}
\usepgfplotslibrary{groupplots}
\pgfplotsset{
    layers/my layer set/.define layer set={
        background,
        main,
        foreground
    }{
    },
    set layers=my layer set,
}
\usetikzlibrary{
    pgfplots.dateplot,
}
\tikzset{
    font={\small\selectfont}
}
\pgfplotsset{compat=newest}

\input defs.tex

\definecolor{green}{rgb}{0, .5, 0}
\definecolor{lightgray}{rgb}{.7, .7, .7}

\newcommand{\hypercube}{
    \draw[->, draw=black] (0,0,0) -- (3,0,0) node[below left] {$x_1$};
    \draw[->, draw=black] (0,0,0) -- (0,3,0) node[right] {$x_2$};
    \draw[->, draw=black] (0,0,0) -- (0,0,3) node[above] {$x_3$};

    \begin{scope}[fill opacity=1, draw opacity=1, text opacity=1]
      \coordinate (p000) at (0, 0, 0);
      \coordinate (p100) at (2, 0, 0);
      \coordinate (p010) at (0, 2, 0);
      \coordinate (p001) at (0, 0, 2);
      \coordinate (p110) at (2, 2, 0);
      \coordinate (p101) at (2, 0, 2);
      \coordinate (p011) at (0, 2, 2);
      \coordinate (p111) at (2, 2, 2);

      \draw [draw=gray] (p000) -- (p100);
      \draw [draw=gray] (p000) -- (p010);
      \draw [draw=gray] (p000) -- (p001);
      \draw [draw=gray] (p100) -- (p110);
      \draw [draw=gray] (p100) -- (p101);
      \draw [draw=gray] (p010) -- (p011);
      \draw [draw=gray] (p010) -- (p110);
      \draw [draw=gray] (p001) -- (p101);
      \draw [draw=gray] (p001) -- (p011);
      \draw [draw=gray] (p110) -- (p111);
      \draw [draw=gray] (p101) -- (p111);
      \draw [draw=gray] (p011) -- (p111);

      \draw [draw=black, fill=black] (p000) circle (1.5pt);
      \draw [draw=black, fill=black] (p100) circle (1.5pt);
      \draw [draw=black, fill=black] (p010) circle (1.5pt);
      \draw [draw=black, fill=black] (p001) circle (1.5pt);
      \draw [draw=black, fill=black] (p110) circle (1.5pt);
      \draw [draw=black, fill=black] (p101) circle (1.5pt);
      \draw [draw=black, fill=black] (p011) circle (1.5pt);
      \draw [draw=black, fill=black] (p111) circle (1.5pt);
    \end{scope}
}

\newcommand{\hypersquare}{
    \draw[->, draw=black] (0,0) -- (3,0) node[below left] {$x_1$};
    \draw[->, draw=black] (0,0) -- (0,3) node[right] {$x_2$};

    \begin{scope}[fill opacity=1, draw opacity=1, text opacity=1]
      \coordinate (p00) at (0, 0);
      \coordinate (p10) at (2, 0);
      \coordinate (p01) at (0, 2);
      \coordinate (p11) at (2, 2);

      \draw [draw=gray] (p00) -- (p10);
      \draw [draw=gray] (p00) -- (p01);
      \draw [draw=gray] (p10) -- (p11);
      \draw [draw=gray] (p01) -- (p11);

      \draw [draw=black, fill=black] (p00) circle (1.5pt);
      \draw [draw=black, fill=black] (p10) circle (1.5pt);
      \draw [draw=black, fill=black] (p01) circle (1.5pt);
      \draw [draw=black, fill=black] (p11) circle (1.5pt);
    \end{scope}
}

\ifjournal
\else
    \fancypagestyle{plain}{%
        \fancyhead{}
        \chead{\textsc{For professional clients and qualified investors only \\
               Not for public distribution \\
               This version posted with permission
               }}
    }
\fi

\title{Portfolio Performance Attribution
\\ via Shapley Value}

\author{Nicholas Moehle \and Stephen Boyd \and Andrew Ang}

\begin{document}
	
\maketitle

\begin{abstract}
We consider an investment process that includes a number of features,
each of which can be active or inactive.  Our goal is to 
attribute or decompose an achieved performance to each of these 
features, plus a baseline value.
There are many ways to do this, which lead to potentially different
attributions in any specific case.  We argue that a specific
attribution method due to Shapley is the preferred method,
and discuss methods that can be used to compute this attribution
exactly, or when that is not practical, approximately.
\end{abstract}

\section{Introduction}

\paragraph{Performance of an investment process.}
We consider an investment process guided by a portfolio manager, 
an automated process, or some combination.  We are interested in some
performance measure of the investment process over some past period, 
for example P\&L, realized return, risk, tracking error, or turnover.
Some of these performance measures we prefer to be large (\eg, P\&L,
realized return), and others we prefer to be small (\eg, risk, tracking error,
turnover).

\paragraph{Features.}
The process has a number of variations or features,
each of which can be active (or on), or inactive (or off).
As a simple example,
consider an investment process that relies
on daily portfolio rebalancing using an optimization method.
The features might be a leverage limit, an ESG constraint that limits
the securities that can be held, and the use of a 
novel return forecast developed by some researchers.  
(In many practical applications, we would have far more than just three features.)

We assume that when the process was run over some time period, all the features 
were active.  In our example above, this means that we ran the process
with the leverage limit, the ESG constraint, and the forecasts.

\paragraph{Attribution.}
The attribution problem asks the question:
How much of the performance should we attribute to each of these features,
and how much to a baseline or benchmark value?
In attribution we are dividing up the performance that was achieved
into an amount associated with each feature, and a baseline value,
which corresponds to what the performance would have been 
with all the features off.
The amount attributed to a feature can be negative, which means that
the feature reduced the performance value.  A positive attribution
means that the feature increased the performance value.
We seek \emph{full attribution}, which means that the sum of the amounts 
attributed to the
features, plus the baseline value, equals the actual performance
achieved.

Continuing our example above, suppose our investment process realizes a
return of 8\% over one year, with all three features active.
An example attribution might be 1\% to the leverage limit, -1\% to the 
ESG constraint, 5\% to the return forecast, and 3\% to baseline performance.
We interpret this as saying that the leverage limit was responsible for
increasing our return by 1\%, while our ESG restrictions depressed the return by 1\%;
our return forecast was responsible for 5\% of the realized return,
and 3\% is attributed to baseline.

Attribution is closely tied to the idea of marginal performance gain,
\ie, the change in performance when a feature is added.
Unfortunately --- and this is the crux of the problem --- the marginal 
performance gain when adding a feature depends on which features
have already been added.
We can think of attribution as assigning a single marginal performance
gain to an feature, independent of which other features are active.

We can carry out attributions for multiple performance objectives.
For example, we can attribute
realized risk, realized return, and realized turnover to our features
and a baseline.  (This is three separate attribution problems.)
An example attribution is shown in table~\ref{t-attr-example}.

\begin{table}
\footnotesize
\centering
\begin{tabular}{ccddd}
\toprule
\multicolumn{1}{c}{$x_1$} & \multicolumn{1}{c}{$x_2$} &
\multicolumn{1}{c}{Risk}  & \multicolumn{1}{c}{Return}  & \multicolumn{1}{c}{Turnover} \\
\midrule
$\times$ & $\times$ & 2.3 & 11 & 43 \\
$\times$ &          & 2   & 12 & 30 \\
         & $\times$ & 1.7 &  8 & 38 \\
         &          & 0.1 &  5 &  2 \\
\bottomrule
\end{tabular}
\hspace{1cm}
\begin{tabular}{ldddd}
\toprule
{} & \multicolumn{1}{c}{Baseline} & \multicolumn{1}{c}{$x_1$}
   & \multicolumn{1}{c}{$x_2$}    & \multicolumn{1}{c}{\emph{Total}} \\
\midrule
Risk      & 0.1 &  1.25 &  0.95 &  2.3 \\
Return    & 5   &  5    &  1    & 11   \\
Turnover  & 2   & 16.5  & 24.5  & 43   \\
\bottomrule
\end{tabular}
\caption{
Example attribution of three performance measures to two
features plus baseline.
The table on the left shows the data,
\ie, the result of backtests with different combinations of the two features.
The table on the right shows an example attribution.
}
\label{t-attr-example}
\end{table}

Attribution has many applications. We can use it to allocate credit, for example
to determine bonus payments.
For features that are associated with an additional cost, we can assess
the cost per unit of performance delivered.
For example if our return forecast in our example incurs a data source cost,
we can compare that cost to the return attributed to it.
We can use attribution to makes changes moving forward; for example,
we can consider dropping features that have a negative attribution to P\&L.
Attribution is a key method for explaining investment performance to clients,
and is often required by law.

\paragraph{Challenges.}
There are two main challenges in carrying out attribution.  The first
is that it involves hypothetical or what-if situations.  While we directly 
observe the performance achieved with all features active, we really do not
know what the performance would have been with some of the features off.
This is addressed by using a high-fidelity simulator or backtester,
that can simulate what would have happened, had some of the features been off.
Of course our attribution can only be as accurate as this simulator.

The second main challenge is conceptual or mathematical.  Except in a very
simple case, which we discuss below, it is hard to exactly define what 
attribution is.
As a result, many different attribution methods are used in practice,
leading to different attributions in any specific case.
In this paper we will argue that a specific type of attribution,
called Shapley attribution, is the best choice. Unfortunately,
with more than just a handful of features, computing the Shapley attribution
exactly requires carrying out an impractically large number of simulations
or backtests. Fortunately, there are methods for computing it approximately,
described below, which work well in practice.

\paragraph{The simple case.}
We describe here, informally, the one case in which attribution is simple. 
(We describe this case mathematically below.)
Attribution is simple when the performance measure is additive, \ie, 
a sum of terms, each associated with a feature plus a constant.
Indeed, this particular form directly gives us the attribution.
As a simple (and uninteresting) example, 
consider the total profit of a company over some
period, with the features being the independent divisions of the company.  
Here the total profit is evidently the sum of the profits contributed by the
divisions, plus some baseline profit for the company that is unrelated to
the profits of the divisions.  

Roughly speaking, in this simple case there is no interaction 
among the features; we simply add up the contributions of the features,
which directly gives us the attribution, with the baseline value 
being the difference 
between the achieved performance and the sum of that attributed to the 
features.

\paragraph{Interactions.}
The challenge is when there are interactions among the features,
in terms of how they affect the performance.
As a simple example, suppose that feature one and feature two,
taken individually,
give no increase in performance; but when they are 
both active, give a substantial improvement in performance.
In this case, how should we attribute the performance to these two 
features?  Intuition suggests they should share the credit equally,
\ie, have attributions equal to half the increase in performance.
As a variation on this situation, suppose that features one
and two are the same, so having them both on is the same as 
having either of them on.  Here too intuition suggests they should 
share the credit.

\subsection{Previous work}

\paragraph{Cooperative game theory.}
Our recommended attribution method uses the Shapley value,
an idea from originated in cooperative game theory,
which attempts to answer the question of how to allocate the earnings
of a coalition to individual players \citep{shapley1953value}.
It is derived axiomatically
by showing that the only attribution method that satifies a set if desirable properties
is the Shapley value.
Because of these desirable properties,
the Shapley value is widely considered to be a fair approach 
to allocating value \citep{moulin2004fair}.
The Shapley value has seen a large number of extensions and variations;
see \citep{monderer2002variations} for a summary.
One important extension assumes certain coalitions of players are disallowed,
which changes the resulting allocation \citep{hiller2018excluded}.

In general, the effort required to compute Shapley values is exponential in the number of players,
which can be prohibitive when the number of players is large.
In this case, the Shapley values can be approximated using Monte Carlo
\citep{castro2009polynomial},
with error bounds given in \citet{maleki2013bounding}.
When the coalition value function has specific properties (\eg, submodularity),
more efficient methods may exist.  (See \citet{liben2012computing}.)


\paragraph{Attribution in machine learning.}
Recently, Shapley values have been used to interpret the output of 
machine learning prediction models,
such as random forests and neural networks
\citep{lipovetsky2001analysis, vstrumbelj2014explaining}.
In this case, the model inputs (or \emph{features}) are modeled as players in a coalition,
and the resulting prediction performance is the value of each coalition.
As a result, mature theory and software exists for approximately computing the
Shapley value for games with many players
\citep{lundberg2017unified}.

\paragraph{Portfolio performance attribution.}
Since the work of \citet{jensen1968performance},
academics have sought to attribute returns of managers
to skill (security selection) vs.\ exposure to rewarded risk factors, like the market portfolio
\citep{fama2010luck, sharpe1992asset}.
Many approaches for performance attribution exist.
The simplest methods break up portfolio return into components,
which sum to the portfolio return
\citep{brinson1986determinants, brinson1991determinants}.
Other standard approaches use the correlation between portfolio weights and returns
\citep{grinold2000active, grinold2006attribution, lo2007alphas}.
These methods are scalable,
and are often used to attribute performance to a large number of predictive signals.

Many of these standard approaches 
use time series of returns or cross-sectional position holdings
which result in unattributed value (or `residuals');
one advantage of the Shapley value is \emph{full attribution},
which means the attributions of individual features
and the baseline sum to the total portfolio return.
In our exposition, we explicitly contrast Shapley attribution to these more widely used techniques.
While Shapley values have been applied to risk decompositions
\citep{ortmann2016link, mussard2008shapley, tarashev2010attributing, colini2018variance},
to our knowledge, ours is the first application of the Shapley value
to the general portfolio performance attribution problem,
and more generally to any statistic that is produced by an investment process.


%
%
%
%
%
%
%


\section{Model}

In this section we fix our notation and describe our model.

\paragraph{Investment process.}
We assume there is an \emph{investment process}
which produces a dynamic (time-varying) portfolio allocation over some time window.
The investment process may depend on market conditions,
the prior portfolio holdings, 
the decisions of analysts or portfolio managers,
and portfolio optimization techniques,
to update the portfolio holdings over time.
In practice, the investment process may be very complicated,
and we intentionally leave the details unspecified.


\paragraph{Features and configuration.}
The investment process has $n$ features that can be (or could have been)
included or excluded, \ie, active or inactive.
These features might represent the choice of a specific benchmark,
choice of sector exposures or asset allocation,
or the contributions of a specific analyst
or signal.
This inclusion or exclusion of feature $i$ is denoted $x_i\in\{0, 1\}$,
with $x_i=0$ meaning that the feature is inactive, and $x_i=1$ meaning
the feature is active.
The collection of these feature status values 
is called a \emph{configuration} of the investment process,
and is represented by the Boolean vector $x = (x_1, \ldots, x_n)$.
For example, $x=(1,0,1,1,0)$ means that features 1, 3, and 4 are active,
while features 2 and 5 are inactive.

We observe that there are $2^n$ possible configurations, which grows rapidly
with the number of features $n$. For $n=10$, there are around 1000 possible 
configurations; for $n=30$, the number is around $10^9$.

\paragraph{The full and zero configurations.}
The configuration $x=(1,1, \ldots, 1) = \ones$ (the vector of all ones)
means that all features are active. 
We refer to this as the \emph{full configuration} or 
\emph{fully featured configuration}.
We will assume that the full configuration is the one that was actually used.
The other $2^n - 1$ configurations are hypothetical; we did not actually
use them.

The configuration $x=(0,0,\ldots, 0)=0$ is called the 
\emph{zero configuration} or the \emph{baseline configuration} or the 
\emph{benchmark configuration}.
It corresponds to the investment process with all features inactive.
In some cases it can be interpreted as investing in a benchmark portfolio.

\paragraph{Performance metric.}
This investment process is evaluated using a real-valued 
performance metric $y\in\reals$.
(In practice, portfolios are evaluated using many metrics,
which can be considered separately.)
Examples of performance metrics
include the portfolio's return, risk, risk-adjusted return,
turnover, or average exposure to a particular risk factor,
over some investment period.
(These can be ex-ante values, evaluated using a contemporaneous model;
or realized, ex-post values, evaluated using the actual data.)
Note that large values of $y$ can be good (as in return),
or bad (as in risk).

We assume that when the full configuration $x = \ones$ was used,
the resulting realized performance was $y^{\rm real}$,
which can be directly observed.
In cases when the baseline or zero configuration can be interpreted
as investing in a benchmark portfolio, the performance value for this too
can be directly observed.

\paragraph{Simulation and backtesting.}
We use simulation to judge the performance using other, hypothetical
configurations.
This typically has the form of a backtesting engine,
which can evaluate the performance under the hypothetical configurations.
This process is represented by a function $f:\{0,1\}^n \to \reals$, with
\[
y = f(x) = f(x_1, \ldots, x_n).
\]
Evaluating the function $f$ requires running a backtest
of the investment process under configuration $x$,
and recording the performance $y$.
We assume that the backtests are calibrated so that
$y^{\rm real} = f(\ones)$, \ie, the backtest simulation result
for the configuration we used agrees with the performance we actually observed.

\paragraph{Lift or marginal contribution.}
We now introduce a natural concept in attribution, which is the 
change in performance when we add one feature.
Suppose the configuration is $x$, with $x_i=0$, \ie, feature $i$ is inactive.
Then $\tilde x = x+ e_i$ (where $e_i$ is the $i$th unit vector) is 
the configuration obtained by turning feature $i$ on.
We define the \emph{lift} or \emph{marginal contribution} 
as the change in performance obtained
by adding feature $i$, \ie,
\[
f(x+e_i) - f(x).
\]
This marginal contribution depends on the particular configuration $x$.
In other words, the lift associated with adding a feature depends on which other
features are active.

\paragraph{Geometric interpretation.}
We can associate the $2^n$ different possible configurations with the
corners of a unit (hyper)cube in $\reals^n$.
We can create a directed graph with configurations as vertices, by having
an edge from configuration $x$ to configuration $\tilde x$ if $\tilde x = x + e_i$ for some $i$.
In words, an edge goes from one configuration to another that 
is obtained by adding one feature.
We note for future use that we can associate with an edge 
from $x$ to $\tilde x = x+e_i$
a marginal performance change $f(\tilde x)-f(x) = f(x+e_i)-f(x)$.
The number of edges is $n2^{n-1}$.

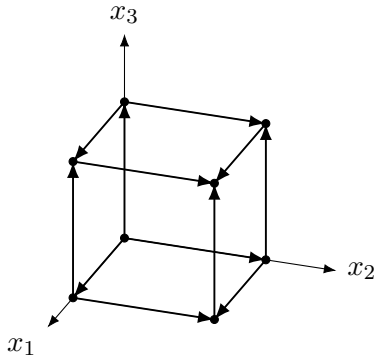
\begin{figure}
\centering
\tdplotsetmaincoords{65}{110}
\begin{tikzpicture}[tdplot_main_coords, >=Latex]
\hypercube
      \draw [->, draw=black, line width=0.25mm] (p000) -- (p100);
      \draw [->, draw=black, line width=0.25mm] (p000) -- (p010);
      \draw [->, draw=black, line width=0.25mm] (p000) -- (p001);
      \draw [->, draw=black, line width=0.25mm] (p100) -- (p110);
      \draw [->, draw=black, line width=0.25mm] (p100) -- (p101);
      \draw [->, draw=black, line width=0.25mm] (p010) -- (p011);
      \draw [->, draw=black, line width=0.25mm] (p010) -- (p110);
      \draw [->, draw=black, line width=0.25mm] (p001) -- (p101);
      \draw [->, draw=black, line width=0.25mm] (p001) -- (p011);
      \draw [->, draw=black, line width=0.25mm] (p110) -- (p111);
      \draw [->, draw=black, line width=0.25mm] (p101) -- (p111);
      \draw [->, draw=black, line width=0.25mm] (p011) -- (p111);
\end{tikzpicture}
\caption{Visualization of configurations as vertices of a hypercube,
for $n=3$.  The vertices, shown as dots, as configurations. The edges
correspond to adding one feature to a configuration.}
\label{f-hypercube}
\end{figure}

An example with $n=3$ features is depicted in figure~\ref{f-hypercube}.
The point $(0, 0, 0)$ represents the baseline configuration,
and the point $(1, 1, 1)$ represents the fully featured configuration.
The edge from $(0,1,0)$ to $(1,1,0)$
corresponds to adding feature 1 to the configuration with only 
feature~2 active.
There are a total of 12 edges.

\section{Attribution}
We would like to attribute the realized performance $y^{\rm real}$ to the $n$ features,
\ie, to determine how much of the performance resulted from each feature.
An attribution method determines real values $a_1, \ldots, a_n$ and $b$,
where $a_i$ is the amount attributed to feature $i$,
and $b$ is the baseline amount.
The attribution and baseline amounts can be positive or negative.
We will represent the attribution using a vector $a = (a_1, \ldots, a_n)$
and scalar $b$.

The attribution is derived from the feature performance function $f$, \ie,
its values for the $2^n$ different configurations.
An attribution method is an algorithm or method that determines the attribution
based on evaluating $f$ for some, or possibly all, configurations.

\subsection{Attribution desiderata}
\label{s-desiderata}
We now describe several desirable properties of an attribution method.

\paragraph{Full attribution.}
We require \emph{full attribution}, which means that
\[
y^{\rm real} = f(\ones) =  a_1 + \cdots +a_n + b = \ones^T a + b.
\]
This means that the observed performance measure $f(\ones)$ is fully 
attributed to the $n$ features, plus the baseline.
Even though full attribution is a crucial property of a good attribution method,
we will see that commonly used attribution methods do not have it.

\paragraph{Correct baseline value.}
We say that an attribution has the correct baseline value if 
\[
f(0) = b.
\] 
This means that the baseline value $b$
matches the performance of the benchmark portfolio.

\paragraph{Fairness.}
We call an attribution method fair if,
for any permutation of the features,
the attributions are permuted the same way.
This property implies that if two features are the same, \ie,
they have the same effect on performance, then their attributions must be the same.

\paragraph{Monotonicity.}
This means that if we change $f$ in such a way that one feature's marginal contribution 
does not decrease (no matter which features are already active),
then the attribution to this feature also does not decrease.

\subsection{Additive case}
We say the performance is \emph{additive} if $f$ has the form
\[
f(x) = a^T x + b = b + \sum_{i \; : \; x_i=1} a_i,
\]
for some vector $a$ and scalar $b$.
(If we consider $x_i$ to be real numbers, and not just $0$ or $1$ as we do here,
this corresponds to $f$ being an affine function.)
When $f$ is additive, the baseline performance is $b$, and 
the marginal increase in the performance when adding feature $i$ is 
always $a_i$, independent of what other features are already active.

For an additive function, $a$ and $b$ directly give an attribution
which satisfies all four of the desiderata listed above:
full attribution, correct baseline, fairness, and monotonicity.
The case of additive performance is the easy, or even trivial case,
for attribution.  It is when $f$ is not additive that it becomes more
difficult to assign an attribution.

\section{Attribution methods}
\label{s-methods}
In this section, we describe several attribution methods,
concluding with Shapley attribution method we recommend.

\subsection{One-at-a-time attribution}
We take $b=f(0)$, and 
\[
a_i = f(e_i)-f(0), \quad i=1, \ldots, n.
\]
In other words, we carry out a baseline simulation (if needed) and $n$ additional
simulations, each with exactly one feature enabled.
We attribute to each feature the increase in performance when it is added
to the baseline configuration, \ie, its lift or marginal performance increase from 
the baseline configuration $x=0$.
This method is natural, and requires carrying out only $n$ 
(or $n+1$ if we include the baseline) simulations.

One-at-a-time attribution satisfies correct baseline value, fairness, and monotonicity.
However, it can (and often does) fail to satisfy full attribution.
To see this, consider the simple example with $n=2$ and 
\begin{equation}
\label{e-counter-example}
f(0)=0, \quad f(e_1)=1, \quad f(e_2)=1, \quad f(\ones)=1.
\end{equation}
In this example the presence of either feature $1$ or feature $2$ gives
the full performance value $1$.
One-at-a-time attribution for this example is $b=0$, $a_1=1$, $a_2=1$, so
$b+a_1+a_2=2$, whereas $f(\ones)=1$. 
Roughly speaking, one-at-a-time attribution over-allocates performance
to the features in this example.

We note that one-at-a-time attribution coincides with the
attribution described above for additive $f$.
Here, however, the same formula is being applied to any $f$, not just additive $f$.


\subsection{Leave-one-out attribution}
The leave-one-out attribution method is closely related to one-at-a-time attribution
As in one-at-a-time attribution, we set $b=f(0)$.
We then carry out $n$ simulations, with $x=\ones-e_i$, $i=1, \ldots, n$.
In other words, for each feature we simulate the performance when it is left out,
but all other features are present.
We set
\[
a_i = f(\ones) - f(\ones-e_i), \quad i=1, \ldots, n,
\]
which is the marginal performance increase when adding feature $i$ when all other
features are active.
Like one-at-a-time attribution, 
leave-one-out attribution requires carrying out $n$ simulations, 
plus a baseline simulation.

Like one-at-a-time, leave-one-out attribution satisfies correct baseline value,
fairness, and monotonicity, but it can fail to achieve full attribution.
To see this, consider the same example describe above in~(\ref{e-counter-example}).
The leave-one-out attribution for this example is $b=0$, $a_1=0$, and $a_2=0$,
\ie, it allocates zero performance to each feature.
Roughly speaking, it under-allocates performance in this example.


\subsection{Sequential attribution}
Another commonly used method is sequential attribution.
We start by evaluating the baseline configuration performance $b=f(0)$. 
We then simulate the configuration $x=e_1$, \ie, we add the first feature.
We continue adding features until we have all features active.
We take
\[
a_i = f(e_1+\cdots + e_i) - f(e_1+ \cdots + e_{i-1}), \quad i=1, \ldots, n,
\]
which is the marginal contribution to performance from feature $i$,
when the features $1, \ldots, i-1$ are active.
Like one-at-time and leave-one-out attribution, this method requires
$n$ simulations, plus a baseline simulation.

Sequential attribution satisfies full attribution, since
\begin{align*}
b & +a_1+\cdots+a_n \\
&= f(0) + \big(f(e_1)-f(0)\big) + \big(f(e_1+e_2)-f(e_1)\big) + \cdots
+\big(f(\ones)-f(\ones - e_{n})\big) \\
&= f(\ones).
\end{align*}
It also satisfies correct baseline and monotonicity.  

But sequential atttribution does not satisfy
fairness, since the attribution obtained depends on the order in
which the features are added.
The same example above given in~(\ref{e-counter-example}) illustrates this.
Sequential attribution gives $b=0$, $a_1=1$, and $a_2=0$; that is,
the first feature gets attributed the full performance and the second gets none.
If we swap the two features, we assign the full performance to feature two.

Sequential attribution is also called off-the-top attribution, described
in a different form.  We first evaluate $f(\ones-e_n)$, 
the performance when feature $n$ is turned off, with all others on.
We set $a_n = f(\ones)-f(\ones-e_n)$, the marginal increase in adding
feature $n$ when all others are active.
We then evaluate $f(\ones-e_n-e_{n-1})$, the performance when features $n$ and $n-1$
are both inactive, with all others active, and set
\[
a_{n-1}=f(\ones-e_n) - f(\ones-e_n-e_{n-1}),
\]
the marginal performance when we add feature $n-1$, when features $1, \ldots, n-2$ active.
This is readily seen to result in the same attribution as sequential attribution.
The only difference is in how we tell the story: In sequential attribution we add the features
in order one at a time, whereas in off-the-top attribution, we are removing features in
reverse order.

\begin{figure}
\centering
\tdplotsetmaincoords{65}{110}
\begin{tikzpicture}[tdplot_main_coords, >=Latex]
\hypercube
      \draw [->, draw=red, line width=0.25mm] (p000) -- (p100);
      \draw [->, draw=blue, line width=0.25mm] (p000) -- (p010);
      \draw [->, draw=red, line width=0.25mm] (p100) -- (p110);
      \draw [->, draw=blue, line width=0.25mm] (p010) -- (p011);
      \draw [->, draw=red, line width=0.25mm] (p110) -- (p111);
      \draw [->, draw=blue, line width=0.25mm] (p011) -- (p111);
\end{tikzpicture}
\caption{Two permutations for sequential attribution.
The red path corresponds the the standard permutation 
$(0, 0, 0)\to(1, 0, 0)\to(1, 1, 0)\to(1, 1, 1)$,
while the blue path corresponds to the permutation
$(0, 0, 0)\to(0, 1, 0)\to(0, 1, 1)\to(1, 1, 1)$.
Under the permuted attribution method,
the lift associated with each edge in the path is the attribution
to the feature added along that edge.
}
\label{f-permuted-sequences}
\end{figure}
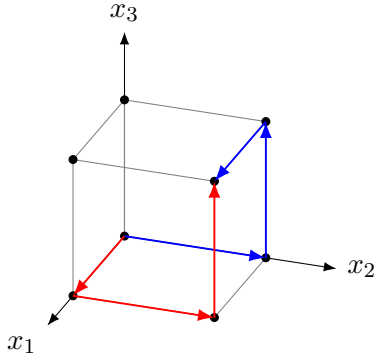

\paragraph{Geometric interpretation.}
We can give a nice geometric interpretation of sequential attribution.
We start at node or vertex $x=0$ and follow a specific directed path to node $e_1$, then $e_2$, 
and so on, ending at the full configuration $x=\ones$.
The attribution to feature $i$ is the marginal increase associated with the 
edge in which feature $i$ is added.
An example with $n = 3$ is shown (in red) in figure~\ref{f-permuted-sequences}.

\subsection{Permuted sequential attribution}
Here we describe a simple extension of sequential attribution that will come up
in the sequel.  Let $\pi = (k_1, \ldots, k_n)$ be a permutation of 
$(1,\ldots, n)$, which means that each integer from $1$ to $n$ appears as one of 
the $k_i$.
Define $\tilde x$ as $\tilde x_i = x_{k_i}$, $i=1, \ldots, n$.
This is the configuration vector when we permute the features using $\pi$.
We define the permuted performance function as $\tilde f(\tilde x) = f(x)$.

Permuted sequential attribution permutes the original features to obtain $\tilde f$,
then uses sequential attribution on $\tilde f$,
and finally permutes the resulting attribution $\tilde a$ and $\tilde b$ back to the original 
ordering.  
In sequential attribution, we
use the marginal performance contribution when the features are added one by one,
in order.  
Permuted sequential attribution is the same, except that we add the features in
the order $(k_1, k_2, \ldots, k_n)$.

Permuted sequential attribution satisfies full attribution, correct baseline,
monotonicity, but not fairness.
Indeed, fairness would require that the attribution obtained is the same
for any permutation $\pi$.
(This is the case if and only if $f$ is additive.)

\paragraph{Geometric interpretation.}
We can associate a permutation $\pi$ with
a directed path from $0$ to $\ones$ on the vertices of the hypercube,
and vice versa, since any such path corresponds to a permutation.
We allocate to each feature the marginal change in performance along the edge in which
feature $i$ is added.
An example with $n = 3$ is shown (in blue) in figure~\ref{f-permuted-sequences}.

\subsection{Shapley attribution}
Finally we come to the attribution method we endorse, the Shapley method.
The Shapley attribution is simply the average of the permuted sequential attributions
over all $n!$ permutations.
Formally, let $a_\pi$ and $b$ denote the attribution for permuted sequential attribution
with permutation $\pi$.  (The value of baseline attribution $b$ does not depend on the 
permutation.)
The Shapley attribution is
\begin{equation}
\label{e-shapley}
a = \frac{1}{n!} \sum_{\pi} a_\pi,
\end{equation}
where the sum is over all $n!$ permutations.

This method satisfies all the desiderata: full attribution, correct baseline, fairness,
and monotonicity.  Indeed, it has been shown that any attribution method that satisfies
these four desiderata must coincide with the Shapley attribution
\citep{young1985monotonic}.

The bad news is that evaluating the Shapley attribution requires 
$2^n$ simulations, which for $n$ larger than
10 or so is likely to be impractical.
This is constrasted with the one-at-a-time, leave-one-out, sequential, and permuted
sequential attribution methods, which require only $n+1$ simulations.
We will address this computational complexity issue in more detail below.

We summarize the properties of the different attribution methods 
in table~\ref{t-methods-desiderata}.
\begin{table}
\small
\centering
\begin{tabular}{lccccc}
\toprule
Method & Full attr. & Baseline & Fairness  & Monotonicity  & Simulations\\
\midrule
One-at-a-time   &            &   $\times$ &  $\times$ &      $\times$  & $n+1$ \\
Leave-one-out   &            &   $\times$ &  $\times$ &      $\times$  & $n+1$ \\
Sequential      &   $\times$ &   $\times$ &           &      $\times$  & $n+1$ \\
Permuted seq.   &   $\times$ &   $\times$ &           &      $\times$ & $n+1$ \\
Shapley         &   $\times$ &   $\times$ &  $\times$ &      $\times$ & $2^n$ \\
\bottomrule
\end{tabular}
\caption{Properties of attribution methods.  Righthand column gives number of 
simulations required to compute the attribution.}
\label{t-methods-desiderata}
\end{table}

\paragraph{Simple example.}
Consider the simple example given in~(\ref{e-counter-example}).
There are only $2!=2$ permutations.
For $\pi=(1,2)$, we get attribution $b=0$, $a_1=1$, and $a_2=0$;
for $\pi=(2,1)$, we get attribution $b=0$, $a_1=0$, and $a_2=1$.
The Shapley attribution for this example is
\[
b=0, \quad a_1 = 1/2, \quad a_2=1/2.
\]
Roughly speaking, in this example, features one and two are the same; the presence 
of either alone gives the full performance.  Permuted sequential attribution gives 
all the credit to the first feature in the sequence, and none to the second.  The Shapley 
attributions average over the two cases, and splits credit to the two features equally.

\paragraph{Geometric interpretation.}
The Shapley attribution for feature $i$ is the 
average of the marginal performance change when feature $i$ is added,
over all directed paths from $0$ to $\ones$.

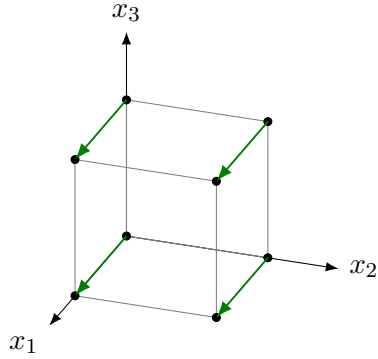
\begin{figure}
\centering
\tdplotsetmaincoords{65}{110}
\begin{tikzpicture}[tdplot_main_coords, >=Latex]
\hypercube
      \draw [->, draw=green, line width=0.25mm] (p000) -- (p100);
      \draw [->, draw=green, line width=0.25mm] (p010) -- (p110);
      \draw [->, draw=green, line width=0.25mm] (p001) -- (p101);
      \draw [->, draw=green, line width=0.25mm] (p011) -- (p111);
\end{tikzpicture}
\caption{
All green edges correspond to adding the first feature to a configuration without it.
}
\label{f-first-feature-lifts}
\end{figure}

\paragraph{The $n=3$ case.} 
We work out the general Shapley attribution for the case with with $n=3$.  
There are $2^n=8$ configurations, and $3!=6$ sequences.
We derive formulas for $a_1$ here; attribution to other features have similar formulas.
First we list the $n!$ sequences and the associated marginal performance change for 
feature $1$ as
\begin{center}
\begin{tabular}{cc}
Permutation & Marginal performance change \\ \midrule
$(1,2,3)$ & $f(1,0,0)-f(0,0,0)$ \\
$(1,3,2)$ & $f(1,0,0)-f(0,0,0)$ \\
$(2,1,3)$ & $f(1,1,0)-f(0,1,0)$ \\
$(2,3,1)$ & $f(1,1,1)-f(0,1,1)$ \\
$(3,1,2)$ & $f(1,0,1)-f(0,0,1)$ \\
$(3,2,1)$ & $f(1,1,1)-f(0,1,1)$.
\end{tabular}
\end{center}
Each line here corresponds to one edge in the graph, in which we add feature $1$.
Some of the marginal changes or edges are repeated; there are only $4$ distinct edges.
Expressing $a_1$ in terms of distinct edges or marginal changes, we have
\begin{align*}
a_1 = \quad \frac{2}{6} &(f(1,0,0)-f(0,0,0))\\ 
\phantom{1} + \frac{1}{6} &(f(1,1,0)-f(0,1,0)) \\ 
\phantom{1} + \frac{2}{6} &(f(1,1,1)-f(0,1,1)) \\ 
\phantom{1} + \frac{1}{6} &(f(1,0,1)-f(0,0,1)).
\end{align*}
The numerators $1$ and $2$ in each line correspond to the number of
paths that include that edge.
For example, there is only one path or permutation that includes the edge from $(0,1,0)$ to
$(1,1,0)$, while there are two paths that include the edge from $(0,0,0)$ to $(1,0,0)$.
In figure~\ref{f-first-feature-lifts}
we show (in green) the four edges corresponding to adding the first feature to a configuration
not containing it.

\section{Computing Shapley attributions}
In this section we focus on methods to compute the Shapley attribution
exactly, or when that is not practical, approximately.
We focus on general methods that work without
any assumptions about $f$.

\subsection{Exact computation}
Computing the Shapley attribution directly using equation~\eqref{e-shapley}
requires taking the average over all $n!$ permuted sequential attributions.
In these sequential attributions,
we end up evaluating $f(x)$ for the same value of $x$ multiple times.
To evaluate the Shapley attribution somewhat more efficiently,
we use an alternative formula for the Shapley attribution,
which sets the baseline value $b = f(0)$
and the attribution to feature $i$ as
\begin{equation}
\label{e-shapley-alt}
a_i = \left(\sum_{x \in\mathcal X_i} \frac{(\ones^T x)! (n - \ones^T x - 1)!}{n!}  \big(f(x + e_i) - f(x)\big) \right) - b.
\end{equation}
Here $\mathcal X_i$ is the set of configurations with feature $i$ off,
\ie, $\mathcal X_i = \{ x \mid x_i = 0\}$.
Using this formula for the
Shapley attributions, it can be computed directly from the 
values of the all $2^n$ configurations.

We note for future use that the coefficients in the 
sum in~(\ref{e-shapley-alt}) sum to one, so they define a probability
distribution on the set of configurations with feature $i$ off.
The formula shows that the $i$th Shapley attribution is the expected value or 
weighted average of the lift obtained by adding feature $i$.

\subsection{Approximate evaluation}
Computing the Shapley attribution requires $2^n$ simulations,
which can be prohibitive when $n$ is large, even just a few tens.
In this case, we recommend approximating the Shapley attributions using 
Monte Carlo sampling methods.  We can either sample over sequences
of lifts, using the two formulas \eqref{e-shapley} and
\eqref{e-shapley-alt}, each of which expresses the Shapley attributions
as an expectation.
The idea of sampling sequences has been proposed in \citet{castro2009polynomial},
but to our knowledge the method based on sampling lifts has not appeared in the literature.

\paragraph{Sampling sequences.}
In this method, we use the sum in definition~\eqref{e-shapley} as a basis for Monte Carlo sampling.
We sample $N$ permutations of the features, with replacement,
and average the permuted sequential attributions corresponding to each.
Computing each permuted sequential attribution requires $n+1$ simulations.
(We can reduce the number of simulations required a bit by caching
previously computed values of $f(x)$, and using these 
when $f(x)$ is needed again.)

The expected value of the attributions corresponding to this method 
are the Shapley attributions.
The Monte Carlo attributions satisfy the full attribution property
(since each permuted sequential attribution does).
The attributions satisfy fairness approximately, or in expectation.

\paragraph{Sampling lifts.}
In this method, we use equation~\eqref{e-shapley-alt} as the basis for Monte Carlo sampling.
As noted above, the sum in \eqref{e-shapley-alt} can be expressed as
\begin{equation}
\label{e-shapley-as-expectation}
a_i = \Expect \big(f(x + e_i) - f(x)\big) - b,
\end{equation}
where the configuration $x$ is random variable
supported over $\mathcal X_i$ with probability distribution
\begin{equation}
\label{e-prob-dist}
\Prob(x = x') = \frac{(\ones^T x')! (n - \ones^T x' - 1)!}{n!}.
\end{equation}

To approximate the Shapley attribution of feature $i$, we first compute $b = f(0)$.
We then sample configurations from $\mathcal X_i$ with distribution~\eqref{e-prob-dist},
and compute the lift of adding feature $i$ to this configuration.
The approximate Shapley attribution is the average of all lifts obtained
minus the baseline value.
This process is then repeated with each feature.
(To sample from distribution~\eqref{e-prob-dist},
first sample the number of active features $\ones^T x$,
which has a multinomial distribution with outcome probabilities $p_i = (i! (n - i - 1)!)/n!$.
Then randomly sample $\ones^T x$ of the $n$ features to be active.)

The advantage of this method is that it tends to produce better approximations with fewer simulations
than by sampling sequences,
because it samples more frequently terms in the sum~\eqref{e-shapley-alt} with larger coefficients,
therefore forming a more precise approximation of the sum quickly.
(This is a form of importance sampling.)
Unfortunately, these approximate Shapley attributions satisfy full attribution 
only in the limit as the number of samples grows, or in expectation.
This can be remedied by scaling the approximate Shapley attributions so that full attribution holds,
even for a finite number of samples.

%

\paragraph{Caching simulations.}
For both sampling methods, some configurations may appear repeatedly.
It is therefore useful to cache the values of configurations,
so they can be re-used in future sampled sequences.
If we are asked to evaluate $f$ for
an $x$ that has already been evaluated,
we simply use the already computed value.

\subsection{Example}
We now demonstrate the approximation techniques for a 
simple numerical example in which the metric is the convex quadratic function
\[
f(x) = x^T P x,
\]
where $P$ is symmetric positive semidefinite.
It can be shown that its Shapley
attribution is $b = 0$ and $a = 2 P \ones$.

We generate $P$ randomly as $P = Z^T Z$,
where the entries of $Z$ are independently drawn from a standard 
normal distribution.
(Thus, $P$ has a Wishart distribution with $n$ degrees of freedom and 
scale matrix $I$.)
We approximate the Shapley attribution using the two methods given above,
sampling sequences and sampling lifts,
and compare the accuracy as a function of the number of 
unique configurations $x$ at which we evaluate $f(x)$.
(This gives the number of configurations for which we evaluate $f$,
using caching as described above.)
When sampling lifts, we consider two versions, 
the basic (unscaled) version and the version in which we
scale so that full attribution holds.
We compare the results using the relative error
\[
e_{\rm rel} = \frac{\|\hat a - a\|_2}{\|a\|_2},
\]
where $a=2 P \ones$ is the true Shapley attribution, and 
$\hat a$ is the sampling-based estimate.

\begin{figure}
    \centering
    \ifmakeplots
        \begin{tikzpicture}
            \begin{axis}[xlabel = Unique configurations evaluated,
                         ylabel = Relative error,
                         xmin=0,
                         xmax=1100,
                         ymin=0,
                         ymax=0.3,
                         table/col sep=comma,
                         width = 1.0\columnwidth,
                         height = 0.5\textwidth,
                         legend cell align={left}
                         ]

                \addplot [thick, blue] table[x=sims, y=rmse]{\datadir/approx_example_sequences.csv};
                \addplot [thick, green] table[x=sims, y=rmse]{\datadir/approx_example_lifts.csv};
                \addplot [thick, red] table[x=sims, y=rmse]{\datadir/approx_example_lifts_scaled.csv};
                \legend{Sampling sequences, Sampling lifts (unscaled), Sampling lifts (scaled)}

            \end{axis}
        \end{tikzpicture}
    \fi
    \caption{
    Relative error of the approximate Shapley attributions 
as a function of the number of unique configurations evaluated,
    when sampling sequences (blue), sampling lifts (green),
    and when sampling lifts and scaling so that full attribution always holds (red).
    }
    \label{f-approx-example}
\end{figure}
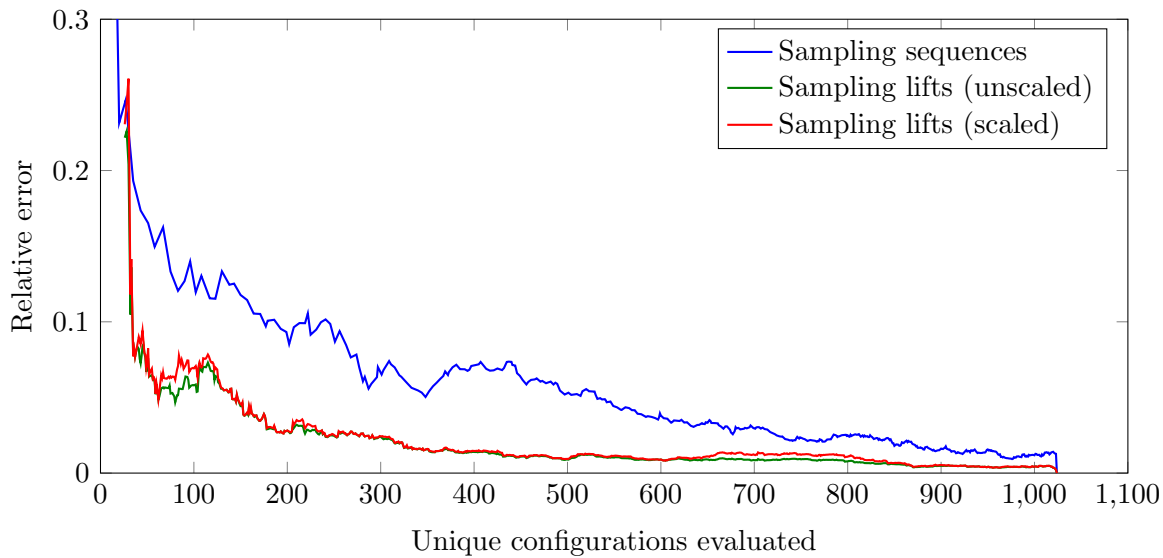

The results are shown in figure~\ref{f-approx-example}, 
for a problem instance with $n=10$.
When sampling sequences,
there is no estimate (and therefore no error) 
until one entire sequence has been evaluated;
similarly, when sampling lifts, there is no estimate or error 
until all features have at least one lift sampled.
Both methods converge to the true values once all $2^{10}=1024$ configurations have been evaluated.
We see that for this example, sampling lifts obtains a lower error
than sampling sequences,
regardless of the number of configurations evaluated.
We also note that when sampling lifts, scaling the approximate Shapley 
values decreases accuracy, but only slightly.
The results of this particular problem instance are typical of many others
we have evaluated.

\section{Examples}

\subsection{Simple return attribution}
\label{s-returns-based-example}
Here we consider a simple return attribution example from \citet[\S 5]{bacon2008practical}.
The performance metric $f(x)$ is the portfolio return over some time period, expressed 
in percent.
We have $n=2$ features:
feature~1 is the country allocation decision
(the decision of how much to invest in which country)
and feature~2 is the stock selection decision
(\ie, the decision of which individual stocks
to hold within these countries).
(When feature~1 is not active, we invest in each country in proportion to its benchmark weight.
When feature~2 is not active, then within each country we invest in each security in proportion to its benchmark weight.)
Table~\ref{t-simple-returns-data} shows example data for a single year.
We revisit this example in appendix~\ref{s-additivity},
where we further decompose returns by country.

\begin{figure}
\centering
\begin{tikzpicture}[>=Latex]
\hypersquare
      \draw [->, draw=blue, line width=0.25mm] (p00) -- (p10);
      \draw [->, draw=blue, line width=0.25mm] (p00) -- (p01);
\end{tikzpicture}
\hspace{1cm}
\begin{tikzpicture}[>=Latex]
\hypersquare
      \draw [->, draw=green, line width=0.25mm] (p00) -- (p10);
      \draw [->, draw=green, line width=0.25mm] (p10) -- (p11);
\end{tikzpicture}
\hspace{1cm}
\begin{tikzpicture}[>=Latex]
\hypersquare
      \draw [->, draw=red, line width=0.25mm] (p00) -- (p10);
      \draw [->, draw=red, line width=0.25mm] (p00) -- (p01);
      \draw [->, draw=red, line width=0.25mm] (p10) -- (p11);
      \draw [->, draw=red, line width=0.25mm] (p01) -- (p11);
\end{tikzpicture}
\caption{
Geometric interpretation of the returns-based attribution example.
The one-at-a-time method attributes based on the lifts
$f(0, 0) - f(1, 0)$ and $f(0, 0) - f(0, 1)$, shown on the left.
The sequential method attributes based on the lifts
$f(0, 0) - f(1, 0)$ and $f(1, 0) - f(1, 1)$, shown in the middle.
The Shapley attribution averages over all possible lifts,
shown on the right.
}
\label{f-returns-attr-example}
\end{figure}
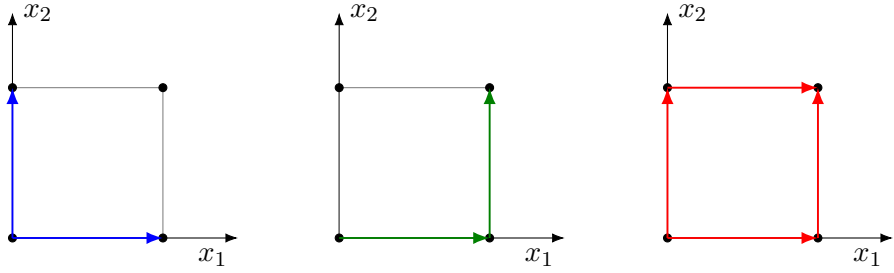

We now discuss, in detail,
how to apply the one-at-a-time, sequential, and Shapley attribution methods to this example,
with a geometric interpretation given in figure~\ref{f-returns-attr-example}.
We will see that the one-at-a-time and sequential methods recover classical attribution methods
known in the literature.

\begin{table}
\centering
\small
\begin{tabular}{cdddd}
\toprule
{}              & \multicolumn{1}{c}{Benchmark}   & \multicolumn{1}{c}{Country alloc.}
                & \multicolumn{1}{c}{Stock sel.}  & \multicolumn{1}{c}{Full portfolio} \\
{}              & \multicolumn{1}{c}{$(0, 0)$}    &  \multicolumn{1}{c}{$(1, 0)$}
                & \multicolumn{1}{c}{$(0, 1)$}    &  \multicolumn{1}{c}{$(1, 1)$} \\
\midrule
$f(x)$     &     6.4   &    5.2   &   9.4    &    8.3 \\
\bottomrule
\end{tabular}
\caption{Data for the simple returns-based attribution example.}
\label{t-simple-returns-data}
\end{table}

\paragraph{One-at-a-time attribution.}
One-at-a-time attribution chooses
\[
b = f(0, 0), \qquad a_1 = f(1, 0) - f(0, 0), \qquad a_2 = f(0, 1) - f(0, 0).
\]
For this example, one-at-a-time attribution is exactly
the classical Brinson--Hood--Beebower method \citep{brinson1986determinants}.
As discussed in section~\ref{s-desiderata},
this method does not have full attribution;
in fact, the unattributed value is $f(1, 1) - f(1, 0) - f(0, 1) + f(0, 0)$.

\paragraph{Sequential attribution.}
Sequential attribution chooses
\[
b = f(0, 0), \qquad a_1 = f(1, 0) - f(0, 0), \qquad a_2 = f(1, 1) - f(1, 0).
\]
This method coincides with the modified Brinson--Hood--Beebower
method given in \citet{bacon2008practical},
where the authors justify choosing feature~1 first in the sequence
because the sector allocation decision is often made before security selection decisions.
This method eliminates the unattributed value,
but violates the fairness property
by prioritizing the sector allocation decision over the stock selection decision during attribution.

\paragraph{Shapley attribution.}
Shapley attribution chooses $b = f(0, 0)$ and
\begin{align*}
\qquad a_1 &= \frac12 \big(f(1, 1) - f(0, 1) + f(1, 0) - f(0, 0)\big), \qquad
\\
a_2 &= \frac12 \big(f(1, 1) - f(1, 0) + f(0, 1) - f(0, 0)\big).
\end{align*}
This is the average of the sequential attributions
produced by the two sequences
$(0, 0)$--$(1, 0)$--$(1, 1)$, which is the sequence used above,
and $(0, 0)$--$(0, 1)$--$(1, 1)$,
which is the sequence obtained by making the stock selection decisions \emph{before}
the sector allocation decisions.

\paragraph{Results.}
Table~\ref{t-simple-returns-results} shows the results of applying all three attribution methods.
As expected, the one-at-a-time method has a non-zero unattributed return.
We can see that in the sequential method,
this unattributed return is entirely allocated to stock selection.
In the Shapley attribution method, the unattributed term is allocated half
to the sector allocation and half to stock selection.
Like the sequential method, it has no unattributed component.
However, unlike the sequential method, it treats country allocation and stock selection equally,
instead of prioritizing country allocation over stock selection.
In this simple and small example,
the differences in attribution by the different methods is not very significant.
In the next section, we will see an example where Shapley attribution is a substantial improvement
over competing methods.

\begin{table}[H]
\centering
\small
\begin{tabular}{ldddd}
\toprule
 & \multicolumn{1}{c}{Benchmark}   & \multicolumn{1}{c}{Country alloc.} 
 & \multicolumn{1}{c}{Stock sel.}  & \multicolumn{1}{c}{Unattributed} \\
 & \multicolumn{1}{c}{$b$}         & \multicolumn{1}{c}{$a_1$}
 & \multicolumn{1}{c}{$a_2$}       & \multicolumn{1}{c}{$y - a_1 - a_2 - b$}      \\
\midrule
One at a time  &   6.4   &   -1.2   &   3.0    &    0.1 \\
Sequential     &   6.4   &   -1.2   &   3.1    &    0   \\
Shapley        &   6.4   &   -1.15  &   3.05   &    0   \\
\bottomrule
\end{tabular}
\caption{Attribution results for the simple returns-based attribution example.}
\label{t-simple-returns-results}
\end{table}


\subsection{Tax-aware portfolio management}
Here we give an example of attribution of multiple performance metrics for a 
tax-aware portfolio management process.
To avoid the wash-sale rule (in which certain capital losses are disallowed),
rebalance trades are carried out monthly.

\paragraph{Metrics.}
We focus on four performance metrics:
realized post-tax return, ex-ante risk, realized capital gains, and 
portfolio turnover.
The return, risk, and turnover are annualized.
The realized capital gains are reported in dollars over the five year simulation.
  

\paragraph{Trading strategy.}
We simulate an investment strategy based on Markowitz portfolio optimization.
In this case, the features correspond to different terms in the optimization problem
that can be on or off.
More specifically, given the configuration
$x$ with $n=7$ features,
we determine the tradelist by solving the optimization problem
\begin{equation}
\label{e-tam-prob}
\begin{array}{ll}
\text{maximize} 
    & \sum_{i=1}^5 x_i h^T \alpha^{(i)} - \gamma \sigma^2(h) - x_6 \ell(h - h_0) - s^T |h - h_0|. \\
\text{subject to}
    & x_7 \sigma(h) \le \sigma_{\rm lim} \\
    & \ones^T h = 1, \quad h \ge 0.
\end{array}
\end{equation}
Here the decision variable is the post-trade portfolio $h$,
expressed as a fraction of the account total;
the pre-trade portfolio (which is given) is $h_0$.
We describe the objective function and constraints in more detail below.

The first term in the objective function is an expected return forecast,
which is divided into the five alpha vectors $\alpha^{(1)}, \dots, \alpha^{(5)}$,
correspending to the momentum, size, quality, value, and minimum volatility factors.
The first five components of $x$ control whether these five alpha vectors are on or off.
The second term is the (scaled) squared active risk, defined as
\[
\sigma^2(h) = (h - h_b)^T \Sigma (h - h_b),
\]
where $\Sigma$ is the return covariance matrix and $h_b$ is the benchmark portfolio,
and $\gamma >0$ is the risk-aversion parameter.
The third term $\ell(h - h_0)$ is the immediate tax liability, due to capital gains,
required to reach the post-trade portfolio $h$,
and is parametrized by the long- and short-term capital gains rates,
and the tax lots comprising the initial portfolio.
(For more details on $\ell$, see \citet[\S 3]{moehle2020tax}.)
This tax-awareness term can be on or off, depending on $x_6$.
The fourth and last term in the objective is a model of transaction cost,
where $s$ is the vector of bid-ask spreads for each asset.

The first constraint is a risk limit with parameter $\sigma_{\rm lim} > 0$.
(When $x_7 = 0$, this constraint is deactivated.)
The second constraint is a full-investment constraint,
and the last constraint specifies that the portfolio is long only.

Note that when $x = 0$, the portfolio aims to simply track the benchmark portfolio.
The full configuration $x=\ones$ means that all seven features are on, 
\ie, we use all five alpha sources, the capital gain objective term, and the risk limit.

\paragraph{Backtests.}
All of our simulations use the S\&P 500 as the benchmark portfolio,
with data over the period 2002 to 2019.
The alpha was obtained using methods similar to those of \citet{kimura2020factors}.
We use the Barra US Equity model \cite{menchero2011barra} to define $\Sigma$ and $h_b$,
and used the risk-aversion parameter $\gamma = 80$.
The tax rates used in $\ell$ were $0.238$ and $0.408$,
which reflect the current highest marginal tax rates in the United States
for long-term and short-term capital gains, respectively.
The simulations take into account transaction costs with the conservative 
value $s = 0.0005 \ones$, \ie, the bid-ask spread is 10 basis points for all assets.
The risk limit is $\sigma_{\rm lim} = 2\%$.

\begin{figure}
\centering 
\pgfplotsset{scaled ticks=false}
\begin{tikzpicture}
\begin{groupplot}[
    group style={group size=1 by 2, vertical sep=1.5cm, horizontal sep=2.5cm},
    yticklabel style={/pgf/number format/fixed},
    ybar,
    table/col sep=comma,
    width = 1\columnwidth,
    height = .5\columnwidth,
    symbolic x coords={baseline, risk_limit, tax_awareness, momentum, minvol, quality, size, value},
    xticklabels={NULL, Base, Risk lim, Tax, Mom., Min.\ vol., Qual., Size, Val.}, 
    cycle list name=mycolorlist,
    after end axis/.append code={
        \draw[thin, color=black] (current axis.left of origin) -- (current axis.right of origin);
    }
]

    \nextgroupplot[
        ylabel=Capital gain (k\$),
        legend to name={CommonLegend2},
        legend style={legend columns=4, /tikz/every even column/.append style={column sep=0.5cm}},
        legend cell align={left},
    ]
        \addplot +[area legend] table[x=index, y=shap]{data/gain.csv};
        \addplot +[area legend] table[x=index, y=oaat]{data/gain.csv};
        \addplot +[area legend] table[x=index, y=loo]{data/gain.csv};

        \addlegendentry{Shapley}
        \addlegendentry{One at a time}
        \addlegendentry{Leave one out}

    \nextgroupplot[
        ylabel=Turnover (\%)
    ]
        \addplot +[area legend] table[x=index, y=shap]{data/turnover.csv};
        \addplot +[area legend] table[x=index, y=oaat]{data/turnover.csv};
        \addplot +[area legend] table[x=index, y=loo]{data/turnover.csv};

\end{groupplot}
\path ($(group c1r2.south) - (0cm, 1.2cm)$) -- node[below]{\ref{CommonLegend2}} ($(group c1r2.south) - (0cm, 1.2cm)$);

\end{tikzpicture}
\pgfplotsset{scaled ticks=true}
\caption{
Attributions of capital gains and turnover for the tax-aware portfolio management example.
}
\label{f-tax-aware-shapley-2}
\end{figure}
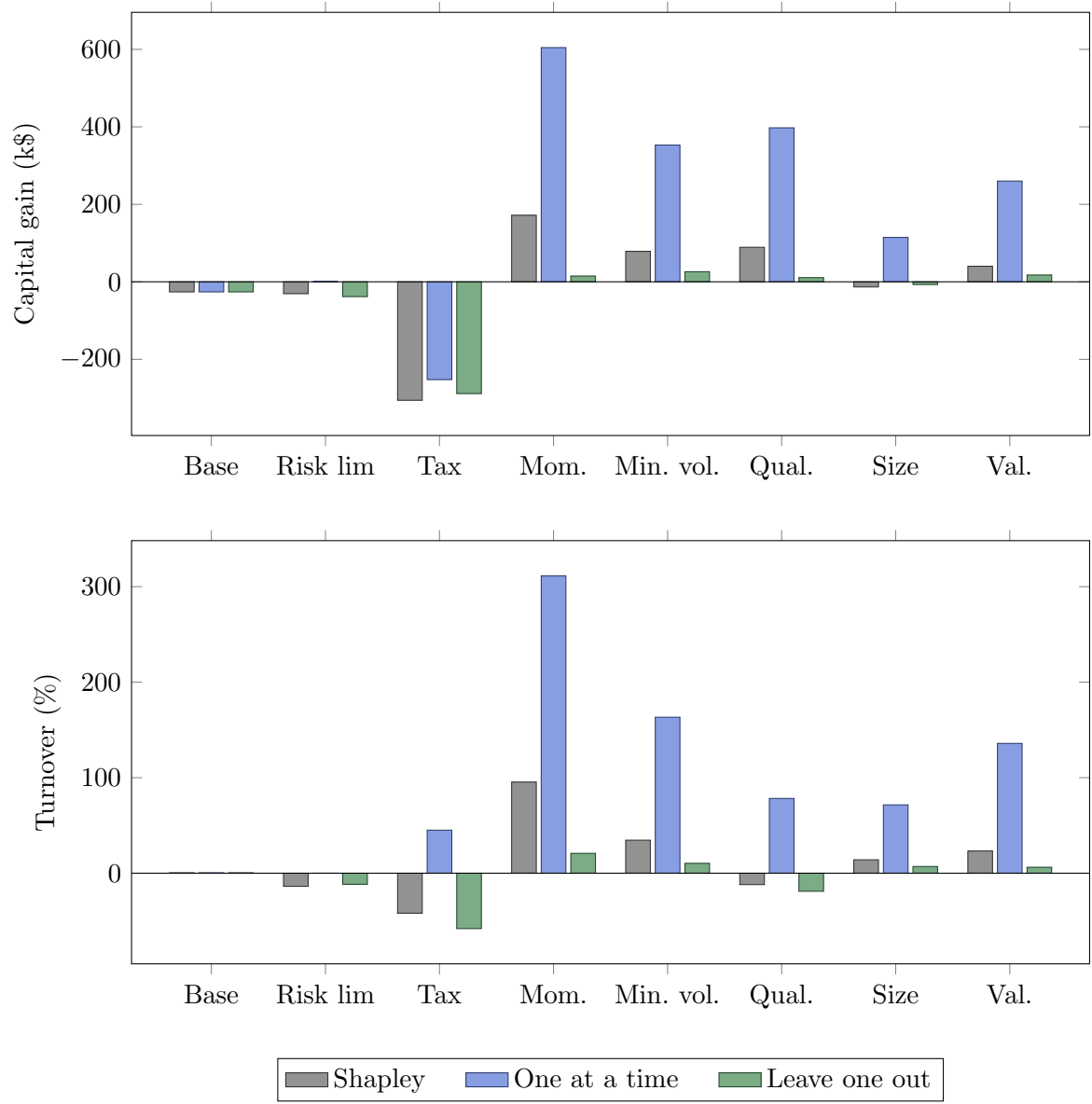

\paragraph{Results.}
Figures~\ref{f-tax-aware-shapley-1} and~\ref{f-tax-aware-shapley-2}
show the attribution results using Shapley, one-at-a-time, and leave-one-out methods.
For each metric, the leftmost set of bars, labeled `Base',
shows the baseline attribution $b$ for each of the three methods.
(The attribution to the baseline is the same for all methods,
as described in section~\ref{s-methods}.)
The following seven sets of bars are the attributions $a_1$, \dots, $a_7$
corresponding to the seven features for each of the three methods.
Table~\ref{t-unattr-example} shows the unattributed amount $f(\ones) - \ones^T a - b$
for each of the three methods and four metrics.
For comparison, we show the metrics for the baseline and full configuration.

By and large, we see the same phenomenon occur for all four metrics:
one-at-a-time attribution over-attributes,
\ie, it overestimates the contribution of each feature,
because when only a single feature is included, it drives the portfolio selection process.
One the other hand,
the leave-one-out attribution under-attributes, \ie, it underestimates the contribution of each feature,
because each single feature makes little difference when competing with the other six.
The degree of over- or under-attribution depends on the specific metric and feature in question.

For example, when attributing the risk,
this leads to serious problems with the one-at-a-time and 
leave-one-out attributions
that are resolved by Shapley attribution.
Under one-at-a-time attribution,
the risk limit does not get any `credit' for risk reduction.
This is because the attribution of risk to the risk limit feature 
is the change in risk by adding it to the benchmark portfolio.
Because the benchmark portfolio already has low risk, the risk limit has no effect.
On the other hand,
each of the five signals, when added to the benchmark portfolio, result in a high risk.
Therefore, with one-at-a-time attribution, risk is severely over-attributed to the five signals.
This problem is also apparent in table~\ref{t-unattr-example};
With the one-at-a-time method, the risk is overattributed,
\ie, the sum of the attributions to the features and baseline is $14.7$\%,
which is is much greater than the true (full configuration) value of $2.0$\%.

Leave-one-out attribution also fails to produce a satisfactory result.
In this case, leaving out one signal while keeping the other four
does not result in a risk reduction at all, due to the active risk limit.
We therefore do not attribute any risk to any of the signals.
In fact, we reach the paradoxical conclusion that even though the benchmark has low risk
and our realized portfolio has high risk,
none of the features are attributed any risk at all.
This problem is again reflected in table~\ref{t-unattr-example}.
With the leave-one-out attribution, the risk is severely underattributed:
the sum of the attributions to the features baseline is $-1.0$\%,
while the true risk was $2.0$\%.

In some cases, the attribution of a metric to a feature has different signs under two different methods.
For example, the tax-awareness feature increases turnover when it is the only feature used,
because it causes the portfolio to realize losses and keep gains,
leading to a positive attribution of turnover to tax awareness under one-at-a-time attribution.
When many other features are active, however, the desire to hold onto tax lots with low basis
means that tax awareness \emph{decreases} turnover;
this is reflected in the negative attribution of turnover to tax awareness
with Shapley and leave-one-out methods.

\newcommand{\baselinecolor}{black}
\newcommand{\risklimcolor}{red}
\newcommand{\taxawarecolor}{green}
\newcommand{\momentumcolor}{magenta}
\newcommand{\minvolcolor}{yellow}
\newcommand{\qualitycolor}{blue}
\newcommand{\sizecolor}{red}
\newcommand{\valuecolor}{red}

\definecolor{color9}{HTML}{636363} 
\definecolor{color1}{HTML}{428953} 
\definecolor{color2}{HTML}{CE2929} 
\definecolor{color3}{HTML}{A3DD57} 
\definecolor{color4}{HTML}{77E599} 
\definecolor{color5}{HTML}{5675D6} 
\definecolor{color6}{HTML}{65ECEF} 
\definecolor{color7}{HTML}{FF8B07} 
\definecolor{color8}{HTML}{D0B100} 

\pgfplotscreateplotcyclelist{mycolorlist}{%
color9!50!black,fill=color9!70!white\\%
color5!50!black,fill=color5!70!white\\%
color1!50!black,fill=color1!70!white\\%
color4!50!black,fill=color4!70!white\\%
color3!50!black,fill=color3!70!white\\%
color6!50!black,fill=color6!70!white\\%
color7!50!black,fill=color7!70!white\\%
color2!50!black,fill=color2!70!white\\%
color8!50!black,fill=color8!70!white\\%
}

\newcommand{\shapdatafile}{data/tax_aware_shapley_data.csv}
\newcommand{\loodatafile}{data/tax_aware_loo_data.csv}
\newcommand{\oaatdatafile}{data/tax_aware_oaat_data.csv}
\newcommand{\taxdatafile}{data/tax_aware_data.csv}

\begin{figure}
\centering 
\pgfplotsset{scaled ticks=false}
\begin{tikzpicture}
\begin{groupplot}[
    group style={group size=1 by 2, vertical sep=1.5cm, horizontal sep=2.5cm},
    yticklabel style={/pgf/number format/fixed},
    ybar,
    table/col sep=comma,
    width = 1\columnwidth,
    height = .5\columnwidth,
    symbolic x coords={baseline, risk_limit, tax_awareness, momentum, minvol, quality, size, value},
    xticklabels={NULL, Base, Risk lim, Tax, Mom., Min.\ vol., Qual., Size, Val.}, 
    cycle list name=mycolorlist,
    after end axis/.append code={
        \draw[thin, color=black] (current axis.left of origin) -- (current axis.right of origin);
    }
]

    \nextgroupplot[
        ylabel=Return (\%),
        legend to name={CommonLegend},
        legend style={legend columns=4, /tikz/every even column/.append style={column sep=0.5cm}},
        legend cell align={left},
    ]
        \addplot +[area legend] table[x=index, y=shap]{data/return.csv};
        \addplot +[area legend] table[x=index, y=oaat]{data/return.csv};
        \addplot +[area legend] table[x=index, y=loo]{data/return.csv};

        \addlegendentry{Shapley}
        \addlegendentry{One at a time}
        \addlegendentry{Leave one out}

     \nextgroupplot[
         ylabel=Risk (\%),
     ]
        \addplot +[area legend] table[x=index, y=shap]{data/risk.csv};
        \addplot +[area legend] table[x=index, y=oaat]{data/risk.csv};
        \addplot +[area legend] table[x=index, y=loo]{data/risk.csv};



\end{groupplot}
\path ($(group c1r2.south) - (0cm, 1.2cm)$) -- node[below]{\ref{CommonLegend}} ($(group c1r2.south) - (0cm, 1.2cm)$);

\end{tikzpicture}
\pgfplotsset{scaled ticks=true}
\caption{
Attributions of return and risk for the tax-aware portfolio management example.
}
\label{f-tax-aware-shapley-1}
\end{figure}
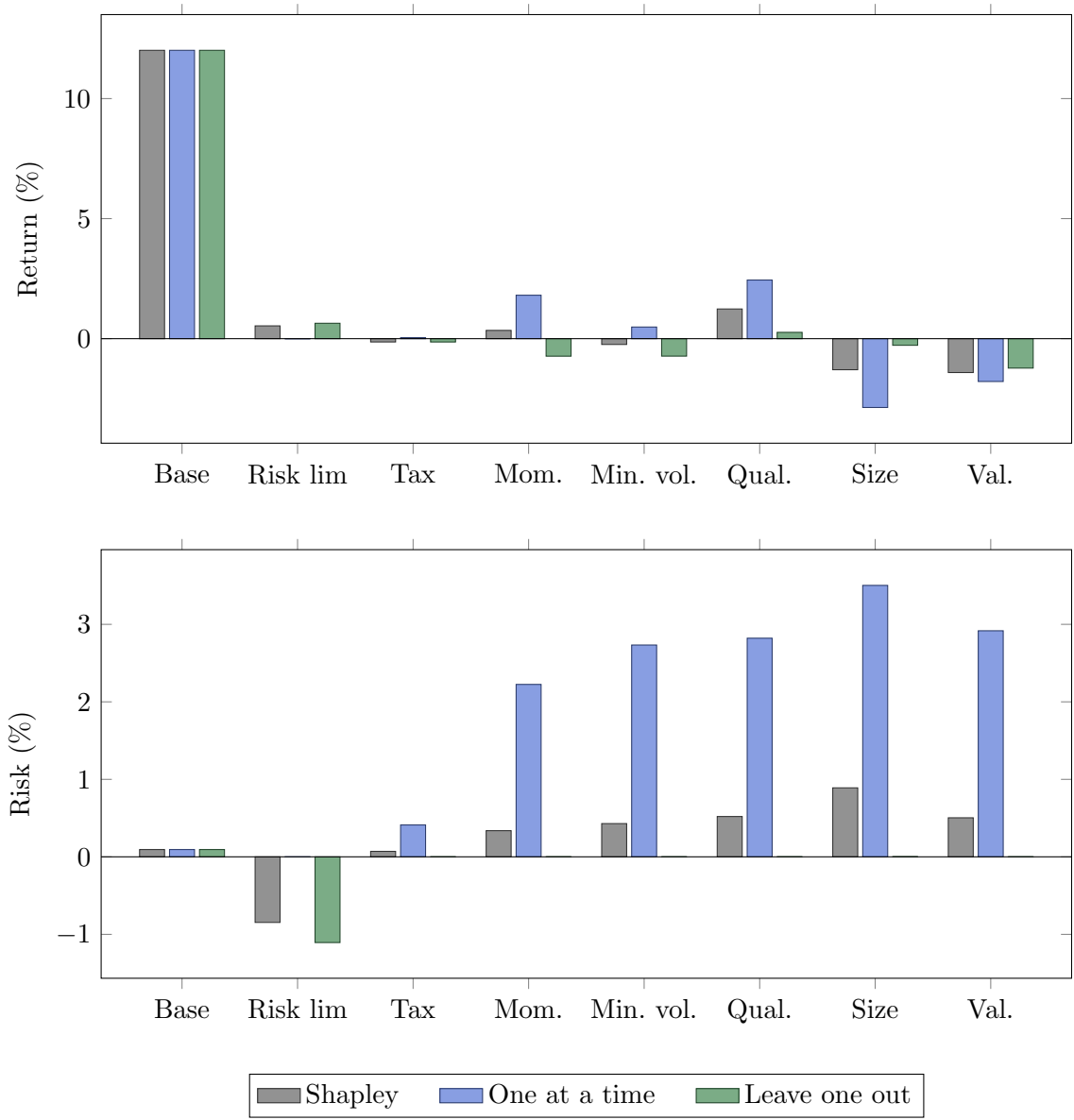

\begin{table}
\footnotesize
\centering
\begin{tabular}{ldddd}
\toprule
{}              & \multicolumn{1}{c}{Return (\%)}      & \multicolumn{1}{c}{Risk (\%)}
                & \multicolumn{1}{c}{Tax (k\$)}         & \multicolumn{1}{c}{Turnover (\%)} \\
\midrule
baseline value $f(0)$             &        12.0   &        0.1 &     -26.0  &     0.0 \\
full configuration value $f(1)$   &        11.0   &        2.0 &       4.6  &   100.2  \\
\midrule
unattributed, Shapley         &         0.0   &      0.0   &       0.0  &     0.0  \\
unattributed, one at a time   &        -1.0   &    -12.7   &   -1447.1  &  -705.6  \\
unattributed, leave one out   &         1.2   &      3.0   &     294.6  &   143.8  \\
\bottomrule
\end{tabular}
\caption{
The baseline value $b = f(0)$, full configuration value $f(\ones)$,
as well as the unattributed components $f(\ones) - \ones^T a - b$
for all three attribution methods.
}
\label{t-unattr-example}
\end{table}

\section{Conclusion}
We propose the use of the Shapley value for portfolio performance attribution.
Shapley attribution is the only method that possesses four properies
that we believe are crucial for satisfactory portfolio performance attribution:
fairness, correct baseline, full attribution, and monotonicity.
(A fifth property, additivity, is discussed in appendix~\ref{s-additivity}.)
We then compare Shapley attribution to other well-known attribution methods.
Compared to other attribution methods,
the only disadvantage of Shapley attribution is computational:
the number of simulations required to carry out Shapley attribution
is exponential in the number of features we attribute to.
To overcome this, we recommend two Monte Carlo methods to approximate the Shapley attribution.
One of these is known from the literature,
and one is novel, to the best of our knowledge.

\paragraph{Acknowledgements.}
We would like to thank Eric Kisslinger for supporting us in carrying out the backtests 
for the tax-aware portfolio management example.
We would also like to thank Ronald Kahn and Isaac Mao for useful
early discussions and testing of Shapley attribution.

\clearpage


\printbibliography

\appendix

\section{Additivity}
\label{s-additivity}
In addition to the desiderata of section~\ref{s-desiderata},
Shapley attribution is \emph{additive}.
This means that if the metric can be decomposed into multiple components,
such that $f(x) = f^{1}(x) + \dots + f^{k}(x)$,
then the Shapley attribution is given by
$a = a^{1} + \dots + a^{k}$ and $b = b^{1} + \dots + b^{k}$,
where $a^{i}$ and $b^{i}$ are the attribution of metric $f^{i}$ to the features.

This is especially useful when the metric is separable across time.
In this case, $f(x)$ is the value of the metric across a large time window (such as a year),
and each $f^{i}(x)$ is the value of the same metric over a shorter time window
(such as a month or quarter).
Examples of time-separable metrics are log-returns and squared risk.

%
%
%

\subsection{Returns-based attribution}
Here we return to the returns-based attribution example from section~\ref{s-returns-based-example},
where we now decompose the returns by country.
Take $f^{\rm uk}(x)$ to be the weighted return on UK stocks,
\ie, it is the portfolio weight in UK stocks multiplied by the return in UK stocks.
(Equivalently, it is the value of UK stocks at the end of the investment period
divided by the initial portolio value.)
Define $f^{\rm jp}(x)$ and $f^{\rm us}(x)$ similarly.
This means that 
\[
f(x) = f^{\rm uk}(x) + f^{\rm jp}(x) + f^{\rm us}(x).
\]

Table~\ref{t-simple-returns-data} shows example data,
which are from \citet{bacon2008practical}.
In particular, the benchmark portfolio weights are 
$40$\% (UK),
$30$\% (Japan),
and
$30$\% (US),
and the portfolio country allociation
was
$40$\%,
$20$\%,
and
$40$\%,
respectively.
The benchmark returns, by country, were 
$10$\%,
$-4$\%,
and 
$8$\ 
respectively,
and the by-sector portfolio returns, after stock selection, were
$20$\%,
$-5$\%,
and 
$6$\%,
respectively.
Combining the data,
we obtain the performance metrics shown in table~\ref{t-sep-returns-data}.

\begin{table}
\centering
\small
\begin{tabular}{cdddd}
\toprule
{}              & \multicolumn{1}{c}{Benchmark}   & \multicolumn{1}{c}{Country alloc.}
                & \multicolumn{1}{c}{Stock sel.}  & \multicolumn{1}{c}{Full portfolio} \\
{}              & \multicolumn{1}{c}{$(0, 0)$}    &  \multicolumn{1}{c}{$(1, 0)$}
                & \multicolumn{1}{c}{$(0, 1)$}    &  \multicolumn{1}{c}{$(1, 1)$} \\
\midrule
$f^{\rm uk}(x)$  &       4   &      4   &     8    &      8 \\
$f^{\rm jp}(x)$  &    -0.8   &   -1.2   &    -1    &   -1.5 \\
$f^{\rm us}(x)$  &     3.2   &    2.4   &   2.4    &    1.8 \\
\hline
Total, $f(x)$    &     6.4   &    5.2   &   9.4    &    8.3 \\
\bottomrule
\end{tabular}
\caption{Data for the returns-based attribution example,
when further sub-divided by country.  }
\label{t-sep-returns-data}
\end{table}

\paragraph{Results.}
In table~\ref{t-sep-returns-results},
we show the results of using the three attribution methods from section~\ref{s-returns-based-example},
but now decomposed by country.

\begin{table}[H]
\centering
\small
\begin{tabular}{lldddd}
\toprule
{} & & \multicolumn{1}{c}{Benchmark}   & \multicolumn{1}{c}{Country alloc.} 
     & \multicolumn{1}{c}{Stock sel.}  & \multicolumn{1}{c}{Unattr.} \\
{} & & \multicolumn{1}{c}{$b$}         & \multicolumn{1}{c}{$a_1$}
     & \multicolumn{1}{c}{$a_2$}       & \multicolumn{1}{c}{$f(x) - a_1 - a_2 - b$}      \\

\midrule
\multirow{4}{*}{One at a time}           & UK            &       4   &      0   &     4    &      0 \\
                                         & Japan         &    -0.8   &   -0.4   &  -0.2    &   -0.1 \\
                                         & US            &     3.2   &   -0.8   &  -0.8    &    0.2 \\
                                         & \emph{Total}  &     6.4   &   -1.2   &   3.0    &    0.1 \\

\midrule
\multirow{4}{*}{Sequential}      & UK            &       4   &      0   &     4    &      0 \\
                                 & Japan         &    -0.8   &   -0.4   &  -0.3    &      0 \\
                                 & US            &     3.2   &   -0.8   &  -0.6    &      0 \\
                                 & \emph{Total}  &     6.4   &   -1.2   &   3.1    &      0 \\


\midrule
\multirow{4}{*}{Shapley}        & UK            &       4   &      0    &     4    &   0 \\
                                & Japan         &    -0.8   &   -0.45   &  -0.25   &   0 \\
                                & US            &     3.2   &   -0.7    &  -0.7    &   0 \\
                                & \emph{Total}  &     6.4   &   -1.15   &   3.05   &   0 \\

\bottomrule
\end{tabular}
\caption{Attribution results for the returns-based attribution example,
when further sub-divided by country.}
\label{t-sep-returns-results}
\end{table}


\end{document}

%% file: defs.tex
\newcommand{\eg}{{\it e.g.}}
\newcommand{\ie}{{\it i.e.}}

\newcommand{\BA}{\begin{array}}
\newcommand{\EA}{\end{array}}

\newcommand{\ones}{\mathbf 1}

\newcommand{\reals}{{\mbox{\bf R}}}

\newcommand{\Expect}{\mathop{\bf E{}}}
\newcommand{\Prob}{\mathop{\bf Prob}}